\documentclass{aastex}

\usepackage{epsf}

\newcommand{\ie}{{\em i.e.}\ }
\newcommand{\eg}{{\em e.g.},\ }
\newcommand{\etal}{{\em et al.}\ }
\newcommand{\cf}{{\em cf.}\ }
\newcommand{\etc}{{\em etc.}\ }

\newcommand{\ngst}{{\em NGST}\ }

\slugcomment{Accepted by the Astronomical Journal}

\shorttitle{Simulated Observations with a Cryogenic Spectrophotometer}
\shortauthors{B.A. Mazin \etal}

\begin{document}

\title{Simulated Extragalactic Observations with a Cryogenic 
Imaging Spectrophotometer}


\author{B.A. Mazin and R.J. Brunner}
\affil{Department of Astronomy, The California Institute of Technology, 
MC 405-47, Pasadena, CA 91125}
\email{bam@astro.caltech.edu,rb@astro.caltech.edu}


%\altaffiltext{1}{}


\begin{abstract}

In this paper we explore the application of cryogenic imaging
spectrophotometers. Prototypes of this new class of detector, such as
superconducting tunnel junctions (STJs) and transition edge sensors
(TESs), currently deliver low resolution imaging spectrophotometry
with high quantum efficiency ($70$--$100$\%) and no read noise over a
wide bandpass in the visible to near-infrared.  In order to
demonstrate their utility and the differences in observing strategy
needed to maximize their scientific return, we present simulated
observations of a deep extragalactic field. Using a simple analytic
technique, we can estimate both the galaxy redshift and spectral type
more accurately than is possible with current broadband techniques.  From our simulated observations and a subsequent discussion of the expected migration path for this new technology, we illustrate the power and promise of these devices.

\end{abstract}




\section{Introduction}

In this paper, we discuss cryogenic imaging spectrophotometers (CIS),
which appear to be the next major revolution in imaging
technology. These devices simultaneously measure the energy, position,
and arrival time of incoming photons with a spectral resolution
$R=\lambda / \Delta \lambda\ $ greater than 20 with high ($>50\%$)
quantum efficiency.  In addition to the improvements in the temporal
resolution over CCDs, these new devices produce low resolution spectra
which are collected without the need for filters, spanning a bandpass
from the atmospheric cutoff at 3100 \AA\ to at least 1 \micron .  The
red limit is imposed by blocking filters designed to keep the rising
sky flux from approaching the maximum count rate of the detectors ---
the devices have an inherent wavelength sensitivity out to around 6
\micron .  As a result, data collected from a cryogenic imaging
spectrophotometer can be collected in ~$20 \%$ of the time as a
conventional CCD and standard filter set (\eg $u$, $g$, $r$, $i$,
$z$), while simultaneously providing at least five times the spectral
information.

The broad band nature of these detectors combined with their spectral
resolution means that techniques designed for CCD photometry or
conventional spectrometers are not optimum for these devices.  For
example, applying a magnitude limit in a specific passband can
eliminate important data from consideration since an object faint in
one passband may contain significant flux at other observed
wavelengths.  The four dimensional data cube returned by these devices
(x position, y position, arrival time, and energy) presents new
challenges in data reduction and analysis, which we address through
simulated observations as discussed in Section 3.

Working CIS detectors for astronomy have recently been demonstrated in
two forms: (1) Superconducting Tunnel Junction detectors (STJs:
\citealt{pf99}), and (2) Transition Edge Sensors (TESs:
\citealt{rm99}).  While these technologies deliver similar data to an
observer, they have their own individual benefits and drawbacks, which
we discuss in more detail in Section 2. Which of these detectors is
expected to dominate in the future (if either --- several promising
new technologies have just emerged) is not clear, and will probably
depend on which detector format is the first to produce an affordable,
large format device.

As a simple example of the analysis of observations of a deep
extragalactic field, we perform a spectral classification of the low
resolution spectra generated by our simulated observations in Section
4. The simple $\chi^2$ minimization approach we utilize, which is very
similar to a template-based photometric redshift determination,
provides a robust estimation for both the galaxy redshift and spectral
type. The significantly higher spectral resolution as compared to CCD
observations (approximately five to ten times the spectral content)
should enable the determination of additional parameters such as age
and star formation rate with more advanced techniques such as a full
eigen-spectra decomposition (\eg ~\citealt{cs00}).

We conclude this paper with a discussion of the implications of this
new detector technology on various areas of Astronomy (\cf
~\citealt{peacock97}), including space-based platforms such as the
Next Generation Space Telescope (\ngst). Throughout this paper we
assume $H_0 = 75$ km s$^{-1}$ Mpc$^{-1}$ and $\Omega_0 = 0.35$ unless
otherwise noted.

\section{Cryogenic Imaging Spectrophotometer Technology}

Cryogenic superconducting detectors hold great promise as general
purpose detectors for ground-based and space-based astronomy.
However, several technical challenges must be addressed to make these
detectors competitive for general use with current instrumentation.

The most severe problem is that current devices have low pixel counts.
Innovative multiplexing solutions will be required to overcome this
problem.  Several multiplexing schemes are currently being tested, and
plans are being made to build large format imaging spectrometers
within the next several years. 

These devices measure individual photons as pulses instead of storing
charge in a potential well.  This allows read noise free output and
accurate timing resolution, but also implies a limited count rate and
a high data rate for a large format array.  A one megapixel device
could produce uncompressed data rates approaching one Gigabyte per
second or multiple Terabytes per night.  Storing and analyzing this
kind of data will present new challenges, and possibly require
complete, on-line reduction.

In a cryogenic spectrophotometer the spectral resolution of the
detectors is dependent on the operating temperature.  Useful devices
start working around $0.3$ Kelvin.  This means expensive He$_3$,
dilution, or adiabatic demagnetization refrigerators will be
necessary.

Another problem is caused by the infrared background.  Since these
devices measure individual photons and are sensitive into the IR,
thermal photons from the telescope and the atmosphere can create a low
level background photon flux which degrades the spectral resolution of the
detectors.  Special blocking filters that thoroughly attenuate the
thermal IR will be needed for these devices to reach their full
potential from the ground.

\subsection{Transition Edge Sensors}

Transition edge sensors work by holding a small piece of
superconductor at its transition temperature by means of
electrothermal feedback.  An incoming photon warms up the detector,
and this causes a decrease in the power needed to maintain the
superconductor at its transition temperature.  This current deficit
pulse is read out by a Superconducting Quantum Interference Device
(SQUID) array.

The TES can ultimately achieve slightly higher spectral resolution
that current generation STJs, but it does have its drawbacks.  The
pixel size is limited by the heat capacity of the
superconductor. Current TES devices are $18 \micron$
square~\citep{rm99}. The SQUID readout, however, cannot be shrunk to
much more that $100 \micron$ on a side.  This means that unless
powerful multiplexing schemes (\eg \citealt{chervenal99}) are
developed a TES will have a low filling factor and need a complex and
expensive readout for each pixel.  The spectral resolution of a TES
depends on its base temperature (currently $\sim~40$ mK), so high
spectral resolution comes at the price of a very complex cooling
apparatus.  Despite these challenges, the TES represent the only
current CIS technology that has the
clear potential to achieve spectral resolution greater than
approximately $60$ at $3000$ \AA.
  
\subsection{Superconducting Tunnel Junction Detectors}

Superconducting tunnel junction detectors work by absorbing a photon
in a strip of superconductor (usually Tantalum) held well below its
transition temperature. The incoming photon breaks up Cooper pairs and
creates free charge carriers called quasiparticles.  The
quasiparticles diffuse through the Tantalum strip and are trapped in
Al-Al$_2$O$_3$-Al Josephson junctions on each end.  The junctions are
connected to either a low noise amplifier or a RF-SET (radio frequency
single electron transistor, see ~\citealt{schoelkopf97}).  Since the
charge created by the photon is distributed over both junctions, the
amount of charge read out in each junction determines the location the
photon hit, and the total charge determines the photon energy.  The
ultimate spectral resolutions of a STJ is limited by the intrinsic
scatter in the number of quasiparticles created by a photon of a given
energy.  This is known as the Fano limit and is a function of the
superconducting energy gap of the material used for the absorber.  For
a Tantalum absorber this works out to approximately $55$ at $3000$
\AA\ and decreases at longer wavelengths as the square root of the
photon energy to $30$ at $10000$ \AA.

STJs have their own unique limitation.  Most designs use back
tunneling in the junctions to increase the signal strength.  This,
along with the shot noise associated with the bias current, lowers the
effective spectral resolution.  The Josephson current in each junction
must be highly suppressed by a magnetic field, thus very uniform
junctions are needed so that a magnetic field will suppress every
junction.

The charge division provided by having two junctions on one absorber
provides STJs with an inherent imaging capability - the number of
virtual pixels on one strip is comparable to the spectral
resolution~\citep{kraus89}. For imaging cameras this means many more
pixels per readout, a real advantage for practical designs.

\section{Simulated Observations}

In order to demonstrate observations with a cryogenic imaging
spectrophotometer, we have generated a simulated extragalactic
field. This operation is complicated by the fact that a CIS
essentially acts as a low resolution spectrograph and we therefore
need not only a simulated galaxy catalog, but also the corresponding
spectra for each source. In the following sections, we detail the
relevant steps required to generate the simulated data, tracing the
photons path from the extragalactic source to the detector. We will
address the more complicated issues of star/galaxy separation,
blending, extended sources, and the identification of active galaxies
in subsequent papers.

\subsection{Galaxy Catalog}

Since we are primarily concerned with demonstrating the utility of CIS
detectors, and not in generating accurate cosmological simulations, we
have used an existing simulated galaxy catalog. In particular, we
started with a galaxy catalog that was drawn from an NGST cosmological
simulation ($\Omega = 0.35$, $H_0 = 75$, Open CDM) performed by
Myungshin Im~\citep{mi00}.  This simulation covers a 2\arcmin\ x
2\arcmin\ area, which, for our purposes, is incidental (we are not
simulating the full three-dimensional data cube, only the individual
sources). This catalog includes the galaxy morphological type,
redshift, and $I$ and $K$ apparent magnitudes.  These parameters were
used to generate spectra for each galaxy in the data set.

\begin{figure}[h]
\epsfxsize=0.5\textwidth \epsfbox{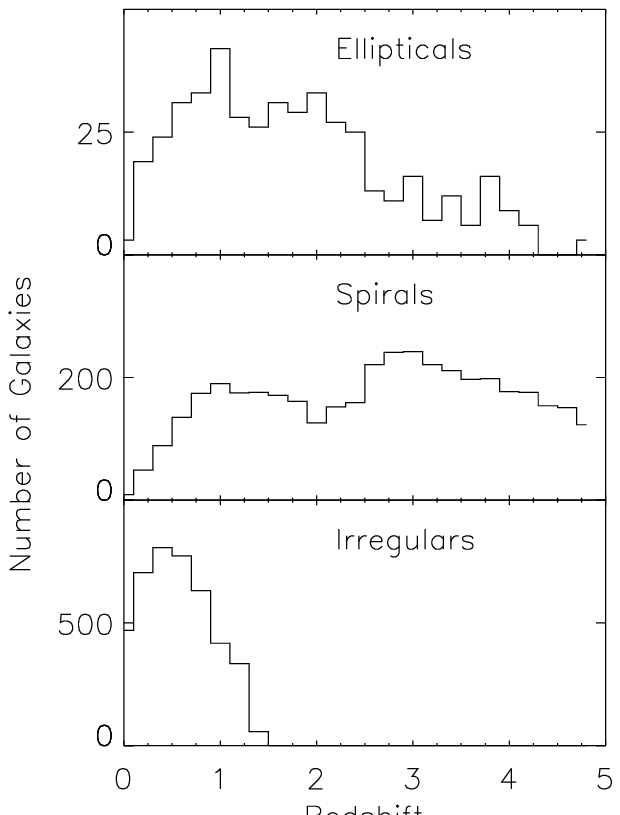}
\caption{Redshift distribution of the three types of galaxies in 
the catalog.
\label{z-dist}}
\end{figure}

We restricted our sample to only include galaxies with $19 \leq I \leq
28$, as galaxies with $I < 19$ are expected to exceed the maximum
count rate of a typical detector, while galaxies with $I > 28$ will
provide an insufficient signal for later analysis.  As a result, our
final galaxy catalog contains $449$ Elliptical galaxies, $4077$ Spiral
galaxies, and $4192$ Irregular galaxies.  The resulting redshift
distributions for each spectral type in the catalog is shown in
Figure~\ref{z-dist}.

\subsection{Model Spectra}

As morphological type is a poorly defined concept at higher redshift,
we interpreted the galaxy types from the original cosmological
simulation as representative of different star formation histories.
We defined the transformation as follows: Ellipticals are simulated by
an instantaneous burst of star formation with subsequent passive
evolution, Spirals are simulated with an exponentially decaying star
formation rate ($\tau = 1$ Gyr), and Irregulars are simulated with a
constant star formation rate.

\begin{figure}
\epsfxsize=0.5\textwidth \epsfbox{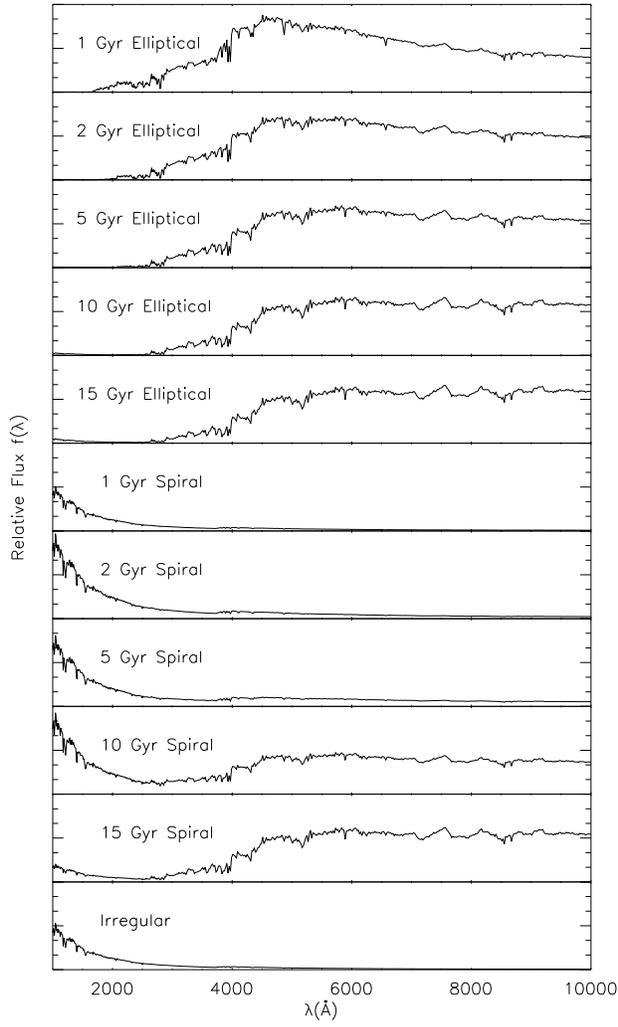}
\caption{The normalized flux of the starting model spectra used to 
simulate the galaxy catalog.
\label{models}}
\end{figure}

The actual model spectra were constructed using the GISSEL
software~\citep{bc98}. The Elliptical and Spiral galaxy models were
generated at ages of 1, 2, 5, 10, and 15 Gyr with solar metallicity
and a Salpeter IMF.  Irregulars were assumed to be inherently young
and only one model was generated.  For each galaxy in the simulated
catalog an age was determined based on its redshift and our assumed
cosmology.  Using this age we selected the closest matching model
spectra. Traces of the complete set of the input model spectra used
can be found in Figure~\ref{models}.

After selecting the appropriate model spectra, we apply various
corrections in order to account for different attenuation factors that
affect the emitted photons. The first modification (which is optional,
in our analysis we simulate observations with and without this
complication) is a correction for cosmological variance (or template
incompleteness).  Template incompleteness results when the template
spectra used to fit the observed data do not span the entire range of
possible galaxy spectra.  This problem is addressed in template-based
photometric redshift estimation by including many templates with
various dust extinction, stellar populations, \etc, which reduces the
contribution of template incompleteness (\eg \citealt{sawicki96}).

Template incompleteness was simulated by adding a single, random
component of stellar light, according to the following likelihood
rule: $25\%$\ O stars, $50\%$\ A stars, $20\%$\ G stars, and $5\%$\ M
stars.  The actual stellar spectra were drawn from a published stellar
library~\citep{pickles98}, scaled by a Gaussian random number with a
sigma of $5\%$\ of the amplitude of the model galaxy spectra, and
added to the original galaxy model spectra.
  
To account for dust in the simulated galaxies, the model spectra were
attenuated using an empirical extinction law~\citep{dc94}. The actual
extinction parameters for the models with cosmic variance were
randomly selected according to a Gaussian probability distribution
function with mean and sigma determined by the following rules:
$E(B-V) = 0.0 \pm .01$ for Ellipticals, $E(B-V) = 0.1 \pm .02$ for
Spirals and $E(B-V) = 0.3 \pm .05$ for Irregulars~\citep{dc94}. The
models without cosmic variance used the mean values described above
with no random component.  In order to avoid unphysical attenuation,
the extinction parameters are restricted to positive values.

\begin{figure}
\epsfxsize=0.5\textwidth \epsfbox{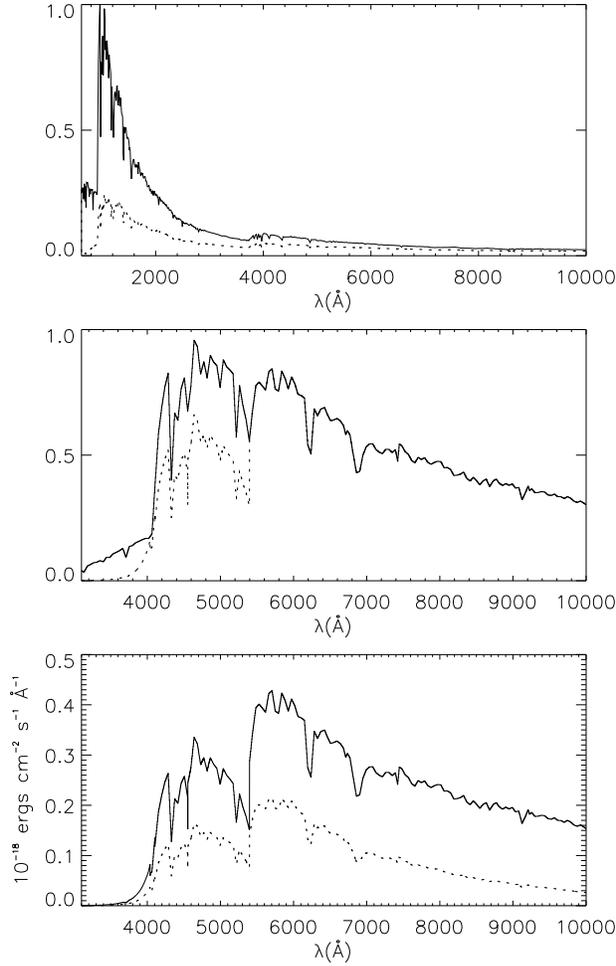}
\caption{Attenuation steps used to generate the spectral energy 
distributions of the input models to the STJ simulator.  A $z=3.44$
spiral in simulated in the above panels.  The top panel contains the
initial~\citet{bc98} model as a solid line.  The dotted line
shows the effect of dust attenuation ($E(B-V)=.1$) in the galaxy.  The solid line in
the center panel shows the redshifted spectra, and the dotted line in
the center panel shows the effects of IGM attenuation.  The solid line
in the bottom panel shows the normalized spectra ($m_{K}=23.88$).  The
dotted line in the bottom panel shows the final model after accounting
for atmospheric attenuation and device quantum efficiency.
\label{steps}}
\end{figure}

Following the dust correction, the spectra were redshifted and
corrected for the mean attenuation due to the inter-galactic
medium~\citep{pm96}.  Each spectrum was normalized by defining the
measured apparent $K$ magnitude to be equal to the original $K$
magnitude from the cosmological simulations.  An SDSS $i'$ magnitude was
then computed for each galaxy.  The final steps in the spectral
simulator are to apply the necessary corrections for the estimated
atmospheric extinction and the combined telescope and device quantum
efficiencies (in this particular simulation we utilize the expected
quantum efficiency for an STJ, see~\citealt{ra97}, and assume a $40\%$\
loss of photons in the telescope and camera optics).  A detailed,
graphical outline of these steps is shown in Figure~\ref{steps}.

\subsection{Broadband Photometric Redshifts~\label{bbpr}}

In order to test the validity of our models and compare them
quantitatively to similar work in the literature we computed broad
band photometric redshifts using our galaxy catalog and associated
model spectra.  This was accomplished by simulating a wide field image
from the Keck telescope with one hour of integration time in each of
four visible bands, $U$,$G$,$R$,$i'$.  For each galaxy a magnitude and
its associated photometric errors were produced.  Using the same
$\chi^2$ template fitting technique described in
Section~\ref{dataanalysis} we fit all galaxies with photometric errors
of less than $0.1^m$ (\ie 10 \% photometry) in each observed band.
The resulting redshift errors, $z_{real}-z_{phot}$ were then used to
determine the standard deviation of the redshift error, $\sigma_{z}$.
In Table~\ref{comparison}, these results are compared to results of
simulations by ~\citet{bj00} using a similar filter set and the
results of ~\citet{lanzetta99} using a combination of HST photometry
and ground based near-infrared photometry.

\begin{deluxetable}{cccccccc}
\tablecaption{A comparison of the rms dispersion in the estimated 
photometric redshifts for template-based photometric redshifts using
space based data~\citep{lanzetta99}, ground-based
simulations~\citep{bj00}, and our no cosmic variance and cosmic
variance simulations. For both the ground-based calculations and our own
simulations, we present the dispersion calculations at both $10$ and
$5$ percent photometric limits.
\label{comparison}}
\tablewidth{0pt}
\tablehead{
\colhead{Redshift Range} &
\colhead{Space Based} &
\multicolumn{2}{c}{Ground Based} &
\multicolumn{2}{c}{No Variance} &
\multicolumn{2}{c}{Variance}
\\
\cline{3-4}
\cline{5-6}
\cline{7-8}
\colhead{} &
\colhead{} &
\colhead{$10 \%$} & \colhead{$5 \%$} &
\colhead{$10 \%$} & \colhead{$5 \%$} &
\colhead{$10 \%$} & \colhead{$5 \%$}
}
\startdata
$0.0 < z < 1.0$	& 0.1	& 0.15  & 0.12  & 0.27	& 0.15    & 0.48	& 0.39 \\
$1.0 < z < 2.0$	& 0.1	& 0.35  & 0.26  & 0.21	& 0.09    & 0.40	& 0.39 \\
$z > 2.0$	& 0.3	& 0.28  & 0.19  & 0.12	& 0.02    & 0.29	& 0.14 \\
\enddata
\end{deluxetable}

It is clear from these simulations that our method of simulating
template incompleteness does in fact broaden the computed redshift
distribution to levels comparable (even slightly higher) than found in
the literature.  The redshift range where we have the greatest
differences with existing broad band data is $z < 1.0$, where our
errors are dominated by irregular galaxies with poor photometry.  As
our set of galaxy spectral templates only includes one Irregular, we
are clearly experiencing the effects of template incompleteness as we
do not properly identify those irregular galaxies with
significant flux from late type stars.

\subsection{Detector Simulations}

\begin{figure}
\epsfxsize=0.5\textwidth \epsfbox{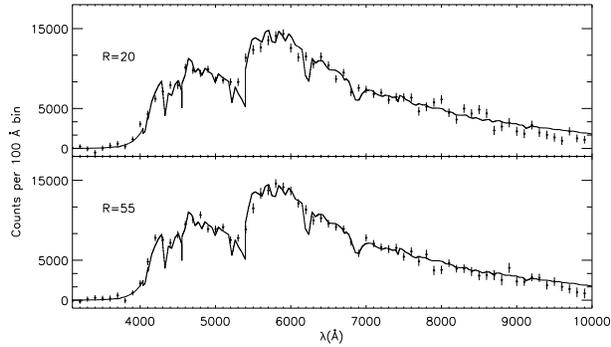}
\caption{Output of the STJ simulator based on the input spectrum from 
Figure~\ref{steps}.  This is a simulated spectra of a z=3.44 spiral
with $m_{K}=23.88$ observed for one hour on Keck with $0.6$\arcsec\
seeing at $R=20$ and $55$.  The solid line is the spectrum computed in
Figure~\ref{steps}.  The dots represent the number of counts in each
100 \AA\ wide bin with their estimated errors.  This galaxy is
detected with a broadband signal to noise ratio of $100$.
\label{sim}}
\end{figure}

Each of the resultant galaxy spectra were pushed through a custom
developed STJ simulator at two spectral resolutions, $R=20$ at $3000$
\AA\ and $R=55$ (Fano-limited) at $3000$ \AA\, with $R$ decreasing as
$1/\sqrt{\lambda}$.  This STJ simulator uses the input spectrum to
perform a photon by photon simulation of the output of the STJ
assuming a one hour integration on the Keck Telescope in $0.6$\arcsec\
seeing. The dominant noise source is sky noise, which is simulated
with the STJ simulator using the sky spectrum of~\citet{bt74}.
Figure~\ref{sim} displays sample output of the STJ simulator.

Our analysis method simulates aperture photometry with a $1$\arcsec\ diameter
circular aperture.  Depending on the seeing conditions and the actual
galaxy diameter, a variable size aperture (\ie a ``total'' magnitude)
could yield a higher signal to noise ratio.  The imaging properties of
the CIS detector will allow observers to dynamically select the best
aperture during data reduction, or to allow spectrophotometry on each
individual pixel.  The latter, for example, could yield interesting
data for extended sources on the differences in age, metallicity, and
dust content of the stellar populations of bulges compared to disks.

\begin{figure}
\epsfxsize=0.5\textwidth \epsfbox{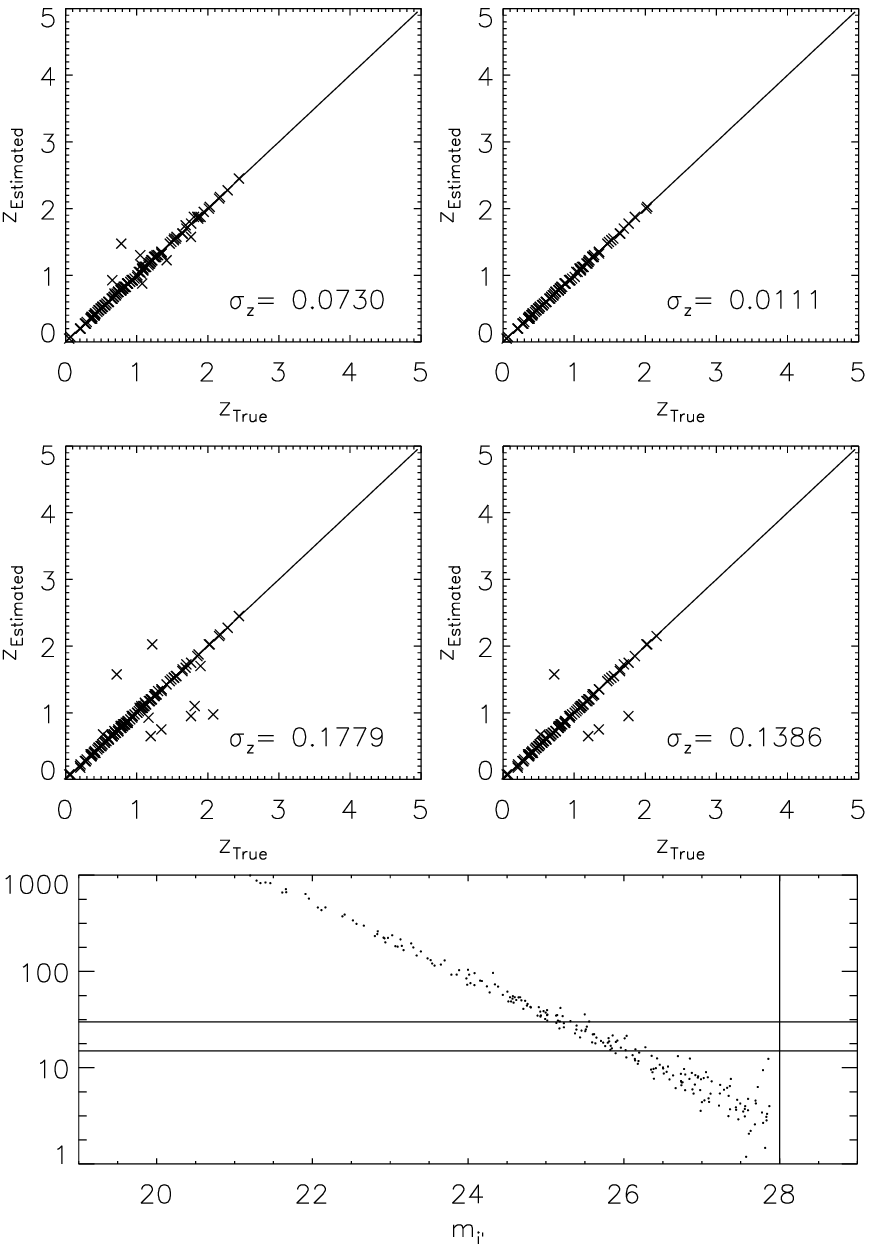}
\caption{The true versus estimated redshift for the simulated Elliptical galaxies 
observed with a $R=55$ STJ detector on Keck for one hour. The top row
contains the simulations without the effects of cosmic variance.  The
second row contains simulations with all of the attenuation factors,
including cosmic variance.  In the first column the fitting algorithm
fits all galaxies with a broadband $S/N$ ratio greater than 15, while
the second column uses a minimum $S/N$ ratio of 30.  The bottom panel
shows the distribution of broad band $S/N$ ratio versus SDSS $i'$
magnitude with the horizontal lines representing the $S/N$ cuts
described previously.
\label{zvz1}}
\end{figure}

\begin{figure}
\epsfxsize=0.5\textwidth \epsfbox{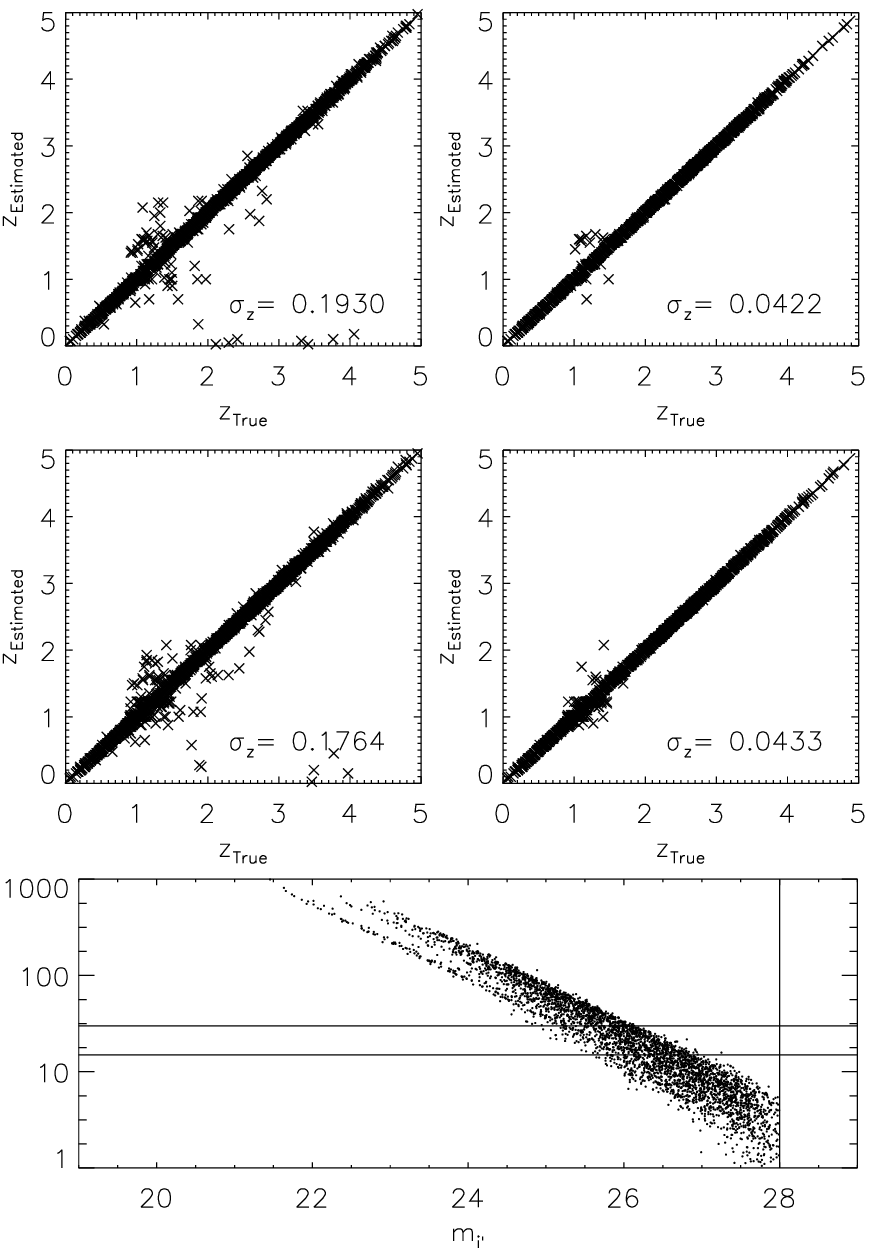}
\caption{The true versus estimated redshift for the simulated Spiral galaxies 
observed with a $R=55$ STJ detector on Keck for one hour. The top row
contains the simulations without the effects of cosmic variance.  The
second row contains simulations with all of the attenuation factors,
including cosmic variance.  In the first column the fitting algorithm
fits all galaxies with a broadband $S/N$ ratio greater than 15, while
the second column uses a minimum $S/N$ ratio of 30.  The bottom panel
shows the distribution of broad band $S/N$ ratio versus SDSS $i'$
magnitude with the horizontal lines representing the $S/N$ cuts
described previously.
\label{zvz2}}
\end{figure}

\begin{figure}
\epsfxsize=0.5\textwidth \epsfbox{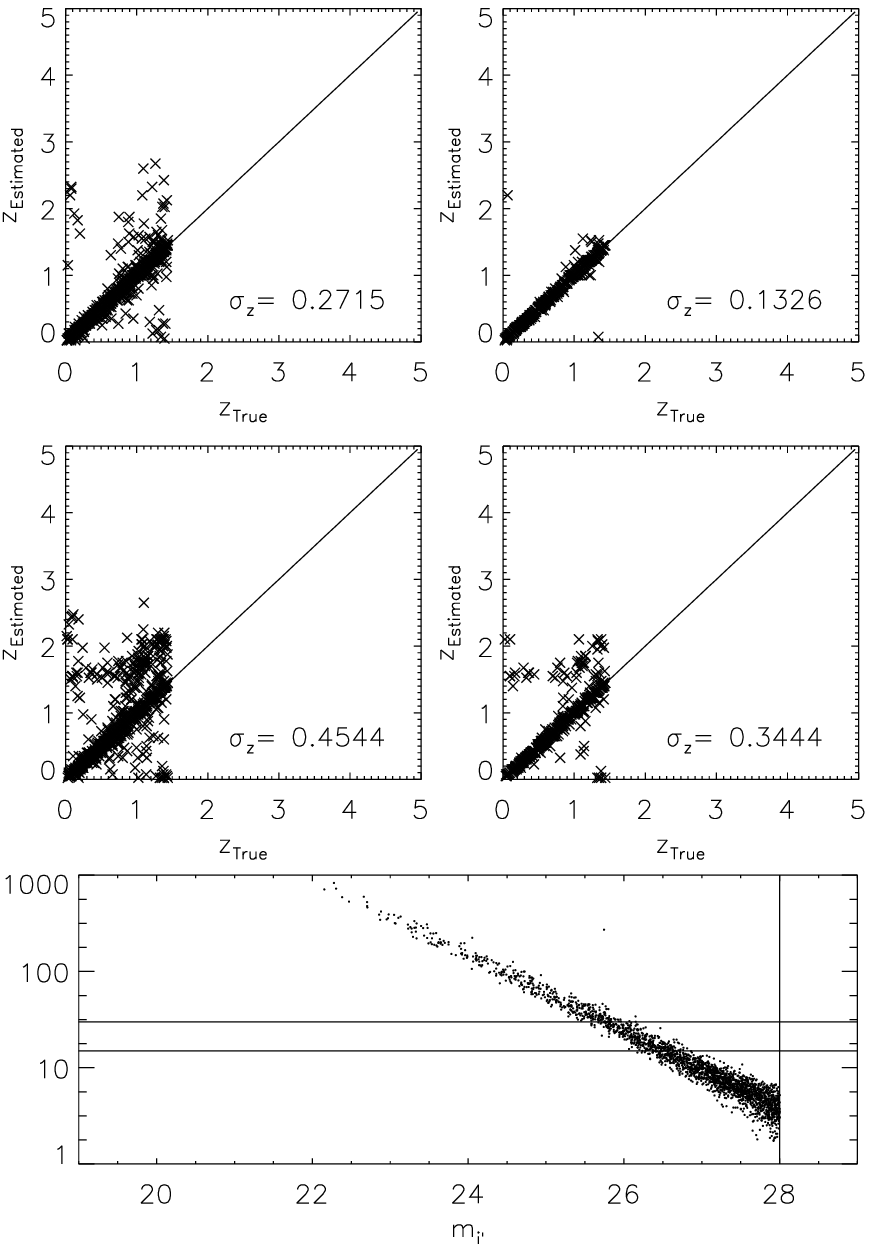}
\caption{The true versus estimated redshift for the simulated Irregular galaxies 
observed with a $R=55$ STJ detector on Keck for one hour. The top row
contains the simulations without the effects of cosmic variance.  The
second row contains simulations with all of the attenuation factors,
including cosmic variance.  In the first column the fitting algorithm
fits all galaxies with a broadband $S/N$ ratio greater than 15, while
the second column uses a minimum $S/N$ ratio of 30.  The bottom panel
shows the distribution of broad band $S/N$ ratio versus SDSS $i'$
magnitude with the horizontal lines representing the $S/N$ cuts
described previously.
\label{zvz3}}
\end{figure}

The broad band nature of these simulated observations makes
conventional selection criteria like an $i'$ band magnitude limit
difficult to apply.  We propose that a correct formalism is to
utilize the broad band signal to noise ratio ($S/N$).  This is just
the total detected source counts divided by the estimated noise
produced by the object plus sky Poisson noise.  This selection
criteria does not translate well into a magnitude limited sample since
an object that is faint in one band may provide significant flux at
other wavelengths that the CIS detector is observing. It does,
however, quantify the low-resolution spectral signal generated by the
CIS detector.  Careful analysis will be needed to quantify the
selection effects introduced by this formalism.  The bottom panel in
Figures~\ref{zvz1},~\ref{zvz2}, and~\ref{zvz3} shows the locations of
the simulated galaxies in a two-dimensional broad band $S/N$ ratio and
SDSS $i'$ band apparent magnitude phase space.

\section{Data Analysis~\label{dataanalysis}}

With the advent of the Hubble Deep Field~\citep{williams96}, the
utility of estimating redshifts from broadband photometry has achieved
wide-spread acknowledgment. Several complementary approaches have been
developed, including techniques which use an empirically derived
function of galaxy flux or colors
\citep{connolly95,brunner97,myThesis,brunner99} and spectral template
techniques which minimize the difference between redshifted empirical
or model templates and the observed photometry to determine redshifts
\citep{gwyn96,lanzetta96,mobasher96,sawicki96}. All of these
techniques, however, are designed to utilize broadband photometry, not
low resolution spectra, and they do not, as a result, extract the
maximal amount of information from our simulated data. Since the major
intent of this paper is not to develop a theoretical framework for
maximal data extraction from CIS observations (which we defer to a
subsequent paper), we instead develop a technique for determining
galaxy parameters which is similar to template photometric redshift
estimation.

In order to determine the redshift of the simulated galaxies, a subset
of the initial model spectra were used.  The following six model
templates were deemed to sufficiently span the template space occupied
by the simulated galaxies: 2 Gyr Elliptical, 10 Gyr Elliptical, 2 Gyr Spiral, 5 Gyr
Spiral, 10 Gyr Spiral, and Irregular.  The model spectra were
attenuated as described in the
previous section to generate a series of template spectra with no
cosmic variance.  A grid of new template spectra were generated by
redshifting these six model spectra over the range $0 \leq z \leq 5$
with $\Delta z = 0.025$.  The new spectra were convolved with a
Gaussian kernel to simulate the effects of an STJ with a spectral
resolution $R=20$ and $R=55$ at $3000$ \AA.
The resulting smoothed spectrum was rebinned into $100$ \AA\ bins to
match the output of the simulator.

\begin{figure}
\epsfxsize=0.5\textwidth \epsfbox{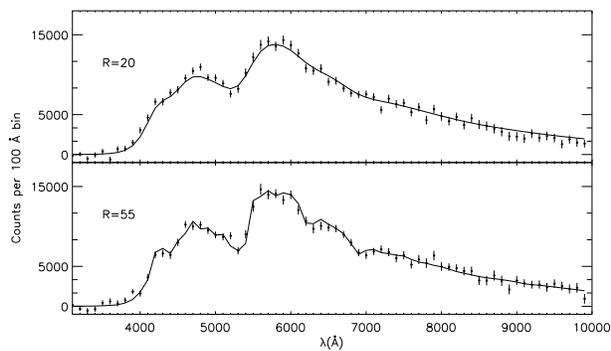}
\caption{The best fit template (2 Gyr spiral at $z = 3.45$) generated by the
redshift estimation algorithm for the simulated STJ spectrum shown in
Figure~\ref{sim}.
\label{simplus}}
\end{figure}

\begin{figure}
\epsfxsize=0.5\textwidth \epsfbox{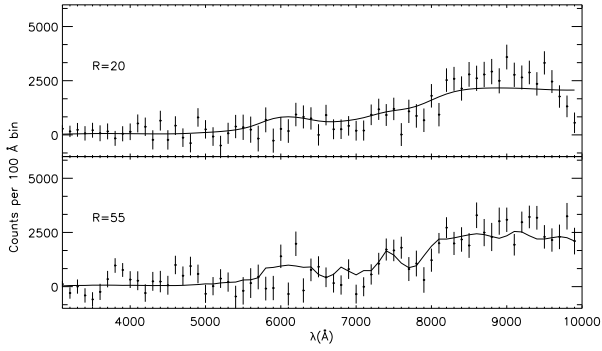}
\caption{A simulated Elliptical at $z = 1.805$ with $m_{K}=21.74$ and 
$m_{i'}=26.05$, detected with a $S/N$ ratio of 17 (which is below our
nominal $S/N$ cut).  The redshift estimation algorithm determined the
best fit template to this galaxy was a 2 Gyr elliptical at $z = 1.725$
which is shown as the solid line.
\label{simplus2}}
\end{figure}

The best fit was determined by finding which template minimized a
$\chi^2$ error statistic~\citep{press92}.  Examples of the fit
returned by this algorithm are shown in Figures~\ref{simplus} and
~\ref{simplus2}.  This method returns both the galaxy spectral
type and redshift.  The measured deviation for all galaxy types is
shown in Figures~\ref{zvz1}, ~\ref{zvz2}, and ~\ref{zvz3}. The results
for the $R=55$ simulations, as well as the lower resolution $R=20$
simulation are summarized in Table~\ref{dispersion}.  Adjusting the
minimum $S/N$ ratio used by the fitting algorithm showed that a $S/N$
ratio of 30 is adequate to provide an accurate redshift.  Lower $S/N$
ratios can cause misidentifications because of noise spikes.

\begin{deluxetable}{cccccccc}
\tablecaption{The dispersion in the estimated redshifts for the different 
simulation parameters (spectral resolution and the inclusion of cosmic
variance) for each of the different star formation histories at
broadband $S/N$ ratios of $15$ and $30$.
\label{dispersion}}
\tablewidth{0pt}
\tablehead{
\colhead{Spectral} &
\colhead{Cosmic} &
\multicolumn{2}{c}{Elliptical} &
\multicolumn{2}{c}{Spiral} &
\multicolumn{2}{c}{Irregular}
\\
\cline{3-4}
\cline{5-6}
\cline{7-8}
\colhead{Resolution} &
\colhead{Variance} &
\colhead{$15$} & \colhead{$30$} &
\colhead{$15$} & \colhead{$30$} &
\colhead{$15$} & \colhead{$30$}
}
\startdata
20 & No & 0.2157 & 0.0474 & 0.1110 & 0.0578 & 0.2929 & 0.1750 \\ 55 &
No & 0.0730 & 0.0111 & 0.1930 & 0.0422 & 0.2715 & 0.1326 \\ 20 & Yes &
0.2923 & 0.2462 & 0.1909 & 0.1008 & 0.4622 & 0.3967 \\ 55 & Yes &
0.1779 & 0.1386 & 0.1764 & 0.0433 & 0.4544 & 0.3444 \\
\enddata
\end{deluxetable}

Clearly, as expected, our simple technique operates most effectively
when there is no cosmic variance in our simulated observations (in
which case we do not suffer from template
incompleteness). Furthermore, as expected, the results for the higher
resolution STJ detector are considerably better than the equivalent
results for the lower resolution detector (see
Table~\ref{dispersion}).

For Elliptical galaxies, our naive approach does a good job at
determining the best redshift and spectral template, except for some
outliers in the simulations which include cosmic variance.  A quick
comparison of the simulations with and without cosmic variance for
those Elliptical galaxies which are detected with $S/N$ above $30$
shows that the small number of outliers increase the dispersion in the
redshift estimation by an order of magnitude.  Upon closer inspection,
these simulated galaxies invariably had significant contributions of
blue stellar light added to their spectrum (similar to an E+A galaxy).
Since there is no comparable template spectrum in our sample (\ie no
E+A template) to model this type of galaxy, our fitting algorithm
misidentifies these cases.  Aside from these rare exceptions, the
dispersion in the estimated redshift versus the simulated redshift for
Ellipticals is lower than for the other galaxy types.  We attribute
this to the greater relative strength of the spectral breaks in the
instantaneous burst models compared to the younger star forming
models.

\begin{figure}
\epsfxsize=1.0\textwidth \epsfbox{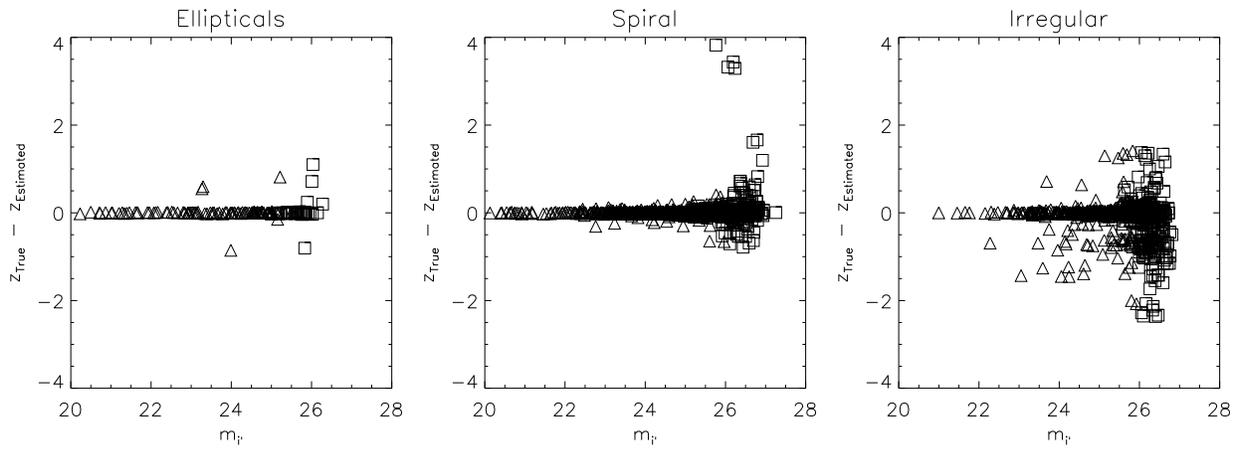}
\caption{The deviation, $z_{True} - z_{Estimated}$, of our simulated galaxies with cosmic variance compared with their SDSS $i'$ magnitude.  Galaxies detected with a $S/N$ between 15 and 30 are represented by boxes.  Galaxies detected with a $S/N$ greater than are 30 represented by triangles.
\label{dzvmi}}
\end{figure}

Another interesting but rare situation arises for a small number of
Irregular galaxies in the cosmic variance simulations, in which the
redshift is systematically misidentified.  This problem is illustrated in Figure~\ref{dzvmi} where high $S/N$ irregulars show a much larger dispersion than other galaxy types at similar magnitudes.  This problem was also evident in the broad band simulations discussed in Section~\ref{bbpr}.  In these special cases, low redshift Irregulars are confused with young $z \backsim 1.5$ Spiral and Irregular templates.  These are mainly detections which contain a significant component of late type stellar light added by the cosmic variance routine.  Since a young Irregular is likely to be dominated by O and B stars, our method for adding cosmic variance can easily overestimate the likely contribution of late type stars.  In a real observation this problem could be corrected by adding a template with a small contribution of
late type stellar light.  This means we expect a more realistic
deviation of Irregulars to be closer to that predicted by the models
with no cosmic variance.

The simulated galaxies in the special cases described above are all
misclassified as a result of template incompleteness.  When the $S/N$
ratio in these galaxies is over about 30, it is possible to determine
which galaxies are effected by template incompleteness by monitoring
the $\chi^2$ goodness of fit parameter~\citep{press92}.  Selecting
and eliminating these galaxies from consideration based on this
parameter would lower the deviation much closer to the simulated
galaxies with no cosmic variance.  However, in a real observation, the
fact that these galaxies do not conform to the standard galaxy models
would make them interesting targets for further study.

One last peculiarity in the measured deviation (somewhat independent
of broadband $S/N$ cut or inclusion of cosmic variance) is the
increase in the dispersion of the relationship for Spiral and
Irregular galaxies with $1 \leq z \leq 2$. This redshift region is a
problem for both photometric and spectroscopic redshift estimation due
to the dearth of spectroscopic features. As a result, galaxies
encounter problems due to template confusion.  A possible solution to
this problem may be obtained by improving our template set, but it
seems likely that the only real solution is to expand the wavelength
coverage of the detectors into the near-infrared.  In the latter scenario,
the 4000 \AA\ break would continue to be sampled on the red side of
the spectrum to sufficiently high redshifts that we would eventually
begin to simultaneously sample the rest frame UV features like the
Lyman break on the blue side of the spectrum.  We discuss several ways
of increasing the wavelength coverage below.

Since the templates were generated on a grid with a spacing of $\Delta
z=0.025$ the maximum error in our redshift estimation due to grid
quantization is 0.0125.  If we assume this error is randomly
distributed, the best we can do is a deviation of 0.0063.  This
systematic error is over $50$\% of our best case dispersion (see
Table~\ref{dispersion}), suggesting that our results can be
significantly improved with more advanced spectral classification
techniques.

Overall, we expect that with real data and a large template set (theoretical or empirical), it will be possible to obtain deviations close to those computed in our simulations with no cosmic variance.  If we ignore the problem area between $z=1.2$ and $z=1.6$ this gives us an expected deviations of less than 0.06 for all galaxy types with $z < 5$.  With the addition of cosmic variance the deviation is broadened to 0.16, which should be considered an upper limit.  These values can be compared with Table~\ref{comparison}, which shows that this one ground based detector can determine redshifts far more accurately than is possible with the best current techniques using Hubble Deep Field photometry and deep Keck IR images. 

\section{Discussion}

These simulations clearly show that cryogenic spectrophotometers
will be far more efficient and accurate than any broad band technique
for redshift determination.  The attraction of these detectors is
obvious.  The minimal five-fold reduction in observing time that is
possible by removing the filters from conventional photometry is
equivalent to transforming the Keck telescope into a 22 meter
telescope for a cost which is considerably lower than building such a
telescope.

From these simulations, it is possible to directly compare the
efficiency of this detector to broadband techniques.  In our most
accurate broadband simulation we identify 1018 out of 8717 galaxies
with the highest detected redshift around 3 using 4 hours of
observation time with Keck.  The CIS detector with $R=55$ and $S/N >
30$ identifies 2045 galaxies much more accurately in only one hour out
to a redshift of 5, translating into an 8 fold efficiency increase.
This advantage, however, is only realized in detectors of the same
size.  Since CCDs currently are very big, CIS detectors will become
competitive for large galaxy redshift surveys when their pixel counts reach
around 1/10 those of CCDs.

We expect that if large format CIS detectors can be built and made
economical they will eventually replace CCDs as the wide field
detectors of choice at most observatories.  The consequences for using
these devices in space is even more dramatic. Due to the removal of
atmospheric effects, space-based CIS devices would allow imaging
spectrophotometry from $1100$ \AA\ to $6$ \micron\ with the same
detector.  A large field imager of this type, mounted on a \ngst
-class telescope with passively cooled mirrors and actively cooled
detectors, would allow images of similar depth to the Hubble Deep
Field with $~2$\%\ of the observation time required with the WFPC2
instrument, and significantly higher spectral coverage and resolution.
Significant performance gains can also be achieved by putting the
ground-based detector behind an adaptive optics (AO) system.  This
would lower the background count rate enough to allow wavelength
coverage out to 2.5 \micron\ with diffraction limited imaging in the
near-infrared.

%  maybe not due to count rate limitations
%Even without AO, further advances in
%data analysis will allow the utilization of the temporal resolution to
%do atmospheric corrections such as a global tip-tilt.

There are many other possible astronomical applications to this
technology.  Applications like an order sorter for an echelle
spectrograph, optical pulsar and X-ray binary work, speckle imaging, interferometry, integral-field spectroscopy, and lensing surveys are
certainly possible and interesting projects, but only scratch the
surface of the potential uses.  In general, almost any project which
uses multicolor photometry on faint sources will be better served by
large format CIS arrays.

The techniques used in this paper for redshift estimation are not
particularly sophisticated.  Indeed, we expect that spectral
classification techniques may prove to be more robust, as well as
extracting more information from the low resolution spectral energy
distributions provided by these detectors.

\section{Conclusions}

In this paper, we have discussed the different types of Cryogenic
Imaging Spectrophotometers which are under development, including
Superconducting Tunnel Junction detectors and Transition Edge
Sensors. We have presented simulated observations of a deep
extragalactic field using a custom developed STJ simulator. While
purposefully ignoring stars and active galaxies in order to focus the
analysis, we have demonstrated that cryogenic spectrophotometers
are an extremely promising new technology for extragalactic astronomy.
Future work in this area will include more detailed simulations of
extragalactic fields including stars and active galaxies, and
observations based from the ground and from space.  More advanced
spectral analysis techniques will also be developed to maximally
extract information from the available data.

\acknowledgments

We are grateful to Pat Cote, Lori Lubin, Brian Kern, Chris Martin, David
Schminiovich, and Todd Small for useful discussions.  More information
on STJ detectors is available at \url{http://spaceball.caltech.edu/}.

%% Generally speaking, only the figure captions, and not the figures
%% themselves, are included in electronic manuscript submissions.
%% Use \figcaption to format your figure captions. They should begin on a
%% new page.

\end{document}